\documentstyle[12pt,psfig,amsmath,amssymb]{article}
\newdimen\singlebaseskip		
\newdimen\doublebaseskip		
\font\eightrm=cmr8			
\font\sc=cmcsc10 scaled\magstep0	

\font\bbbrm=cmbx10 scaled\magstep1
\font\bbrm=cmbx10

\font\tenbmit=cmmib10		
\font\sevenbmit=cmmib10 at 7pt	
\font\fivebmit=cmmib10 at 5pt	
\mathchardef\BOLDalpha="710B 	
 \mathchardef\BOLDbeta="710C 	
\mathchardef\BOLDgamma="710D 	
  \mathchardef\BOLDrho="711A 	
\def\={\overline}		
\def\cms{\ifmmode {\rm\,cm\,\,s^{-1}} \else {$\cms$} \fi} 
\def\cmss{\ifmmode {\rm\,cm\,\,s^{-2}} \else {$\cmss$} \fi} 
\def\deg{\ifmmode {^\circ} \else {${}^\circ$} \fi}	
\def\ergg{\ifmmode {\rm\,erg\,\,g^{-1}} \else {$\ergg$} \fi} 
\def\etal{{\it et al. \/}}	
\def\gcms{\ifmmode {\rm\,g\,\,cm^{-2}} \else {$\gcms$} \fi} 
\def\gcmc{\ifmmode {\rm\,g\,\,cm^{-3}} \else {$\gcmc$} \fi} 
\def\kms{\ifmmode {\rm\,km\,\,s^{-1}} \else {$\kms$} \fi} 
		% ...
\def\lsun{\ifmmode {\rm\,L_\odot} \else {$\lsun$} \fi}	
\def\msun{\ifmmode {\rm\,M_\odot} \else {$\msun$} \fi}	
\def\rsun{\ifmmode {\rm\,R_\odot} \else {$\rsun$} \fi}	
\def\s{\ifmmode \widetilde \else \~\fi} 
\def\spose#1{\hbox to 0pt{#1\hss}} %
\def\Dt{\spose{\raise 1.5ex\hbox{\hskip3pt$\mathchar"201$}}}    
\def\dt{\spose{\raise 1.0ex\hbox{\hskip2pt$\mathchar"201$}}}    
\def\gta{\mathrel{\spose{\lower 3pt\hbox{$\mathchar"218$}}
     \raise 2.0pt\hbox{$\mathchar"13E$}}}
\def\lta{\mathrel{\spose{\lower 3pt\hbox{$\mathchar"218$}}
     \raise 2.0pt\hbox{$\mathchar"13C$}}}

\newcount\notenumber\notenumber=1
\def\foot#1{\raise3pt\hbox{\eightrm \the\notenumber}
     \hfil\par\vskip3pt\hrule\vskip6pt
     \noindent\raise3pt\hbox{\eightrm \the\notenumber}
     #1\par\vskip6pt\hrule\vskip3pt\noindent\global\advance\notenumber by 1}
\def\note#1{\footnote{$^{\the\notenumber}$}{#1}\global\advance\notenumber by 1}
\def\alph#1{\ifcase#1\or a\or b\or c\or d\or e\or f\or g\or h\or i\or j\or
	k\or l\or m\or n\or o\or p\or q\or s\or t\or u\or v\or w\or x\or
	y\or z\else #1\fi}
\def\Alph#1{\ifcase#1\or A\or B\or C\or D\or E\or F\or G\or H\or I\or J\or
	K\or L\or M\or N\or O\or P\or Q\or S\or T\or U\or V\or W\or X\or
	Y\or Z\else #1\fi}

\def\Roman#1{\expandafter\uppercase\expandafter{\romannumeral #1}}
\newcount\sectno    \sectno=0
\newcount\subno     \subno=0
\newcount\subsubno  \subsubno=0
\newcount\eqnmbrsec \eqnmbrsec=0
\newcount\eqnmbr    \eqnmbr=0
\def\sectionbeg#1{\penalty-200\bigskip\par 
	\advance\sectno by 1
	\subno=0\subsubno=0\eqnmbrsec=0
	\noindent{\bbbrm \Roman\sectno. #1}
	\nobreak\smallskip\par}

\def\subsectbeg#1{\penalty-200\medskip\par
	\advance\subno by 1\subsubno=0
	\noindent\hskip 0.25truein {\bbrm \Alph\subno. #1}
	\nobreak\smallskip\par}
\def\subsubbeg#1{\penalty-200\medskip\par
	\advance\subsubno by 1
	\noindent\hskip 0.50truein {\it \number\subsubno. #1}
	\nobreak\smallskip\par}

\def\eqnam#1{\xdef#1{\the\eqnmbr}}		  
\def\eqnamA#1{\xdef#1{{\it A}\the\eqnmbr}}		  
\def\eqnumrun{\global\advance\eqnmbr by 1 \the\eqnmbr}
\def\eqnumsec{\global\advance\eqnmbrsec by 1 \the\sectno.\the\eqnmbrsec}
\def\refindent{\par\noindent\parskip=4pt\hangindent=3pc\hangafter=1 }
\def\refpaper#1#2#3#4#5{\refindent{\sc#1}. #2. {\it #3\/} {\bf#4}, #5.}
\def\refbook#1#2{\refindent{\sc#1}. #2.}
\def\aj#1#2#3#4{\refpaper{#1}{#2}{Astron. J.}{#3}{#4}}

\def\apj#1#2#3#4{\refpaper{#1}{#2}{Astrophys. J.}{#3}{#4}}

\def\icarus#1#2#3#4{\refpaper{#1}{#2}{Icarus}{#3}{#4}}

\def\nature#1#2#3#4{\refpaper{#1}{#2}{Nature}{#3}{#4}}

\def\rmp#1#2#3#4{\refpaper{#1}{#2}{Rev. Modern Phys.}{#3}{#4}}

\def\refpaper#1#2#3#4#5{\refindent{\sc#1}. {\it #3\/} {\bf#4}, #5.}

\def\refpaper#1#2#3#4#5{\refindent{\sc#1}. #2. {\it #3\/} {\bf#4}, #5.}

\begin{document}
 
\begin{center}
\begin{large}
{\bf Jupiter's Obliquity and a Long-lived Circumplanetary Disk} \\
\end{large}
\vskip 0.1 in
Ignacio Mosqueira$^{1,2}$ and Paul Estrada$^1$ 

$^1$ NASA Ames Research Center, $^2$ SETI Institute\\
\end{center} 

\vskip 0.05 in
\begin{abstract}
It has been claimed (Canup and Ward 2002; Ward 2003) that a long-lived
massive (compared to the mass of the Galilean satellites) circumplanetary
gas disk is inconsistent with Jupiter's low obliquity. Such a constraint
could be downplayed on the basis that it deals with a single 
observation. Here we argue that this argument is flawed because it
assumes a solar system much like that of the present day with the one
exception of a circumjovian disk which is then allowed to dissipate
on a long timescale ($10^6-10^7$ yrs). Given that
the sequence of events in solar-system history that fit known
constraints is non-unique, we choose for the sake of clarity of exposition
the orbital architecture
framework of Tsiganis \etal (2005), in which
Jupiter and Saturn were once in closer, less inclined
orbits than they are at present, and show that Jupiter's
low obliquity is consistent with the SEMM
(solids-enhanced minimum mass) satellite
formation model of Mosqueira and Estrada (2003a,b). 

\end{abstract}

\section{Introduction}
\label{sec:intro}

While there may be good reason to believe that planet-disk interactions
may excite the eccentricities of at least isolated planets (Goldreich
and Sari 2003; Ogilvie and Lubow 2003) 
there are currently no studies to guide our understanding
of likely outcomes in the case of
multiple planet (or satellite) systems. Therefore, one 
may consider circular, coplanar giant planets as a starting condition.
Tsiganis \etal (2005) have recently argued that evolution through the
1:2 Jupiter-Saturn mean motion resonance (MMR) 
of a quasi-circular, coplanar and
compact solar system that is allowed to evolve by planetesimal scattering
is consistent with the observed semi-major axes, eccentricities and
mutual inclinations of Jupiter, Saturn, Uranus and Neptune. In a companion
publication, Gomes \etal (2005) argue that this resonance crossing
may be linked to the Late Heavy Bombardment of the terrestrial
planets taking place $\sim 700$ Myr after their formation (Hartmann \etal 2000).
For such a scenario to work 
the bulk of the divergent migration between
Jupiter and Saturn and the resonance passage itself
must take place following gas dissipation, and also 
later than simulations with an evenly spread disk of planetesimals
would indicate. Whether or not the identification with the Late Heavy
Bombardment is correct, in this viewpoint the solar
system must have been more compact and regular prior to
gas dissipation (during its first $10^6-10^7$ years) than it is today. The
authors also briefly consider the potentially disruptive consequences
of such a scenario
for the regular and irregular satellites of the giant planets, but
conclude that at least the regular satellites might have ``survived'' 
unscathed.\footnote{Though the issue is not spelled out in detail (presumably
due to space limitations), in the case of the
Saturnian satellite system Iapetus' relatively low eccentricity
$e \sim 0.03$ may be the main constraint for such a scenario.
For this reason, it would appear unlikely that Titan could owe its
eccentricity to a close encounter between Saturn and another giant planet.}

Jupiter's low obliquity $\sim 3\deg$ may be indicative of its formation
by hydrodynamic gas accretion. Yet, it has been noted that secular
spin-orbit resonances can complicate this straightforward interpretation.
In particular, adiabatic passage through a resonance matching the
spin axis precession rate to the $\nu_{16}$ precession frequency of the orbit
plane due to the gravitational perturbation of
Saturn (Hamilton and Ward 2002; Canup and Ward 2002; Ward 2003)
and $\nu_{17}$ due to the gravitational perturbation of Uranus
(Hamilton, pers. comm.) may result in obliquities
significantly larger than observed. For Jupiter the amplitude of 
the $\nu_{16}$ term (with a period of $\sim 50,000$ years) is
$\sim 0\deg.36$ and the $\nu_{17}$ term (with a period of $P_{16}
\sim 450,000$ yrs)
is $\sim 0\deg.055$, which could have resulted 
in obliquities of up to $\sim 26\deg$ and
$\sim 14\deg$, respectively, had these resonances been crossed
adiabatically as the circumplanetary gas disk was dissipated (Ward 2003).
This leads Canup and Ward (2002) to argue that the circumplanetary
disk must have viscously evolved in a timescale sufficiently short ($O(10^5)$
yr) as to preclude adiabatic passage, resulting in a gas-starved (or
gas-poor [Mosqueira \etal 2000; Estrada and Mosqueira 2005]) satellite disk.

The issue we tackle here is whether our decaying
turbulence\footnote{Consistent
with numerical simulations that show
turbulence decay in the absence of a ``stirring'' mechanism
(Hawley \etal 1999).} SEMM model (Mosqueira and Estrada 2003a,b; hereafter
MEa,b) is especially 
susceptible to secular spin-orbit resonances, and
inconsistent with Jupiter's low obliquity. In particular, we focus
on the $\nu_{16}$ term. The reasons for this are:
First, the period of the orbital precession $\nu_{17}$ ($P_{17} \sim
450,000$ yrs) is already very
close to the precession period of Jupiter's spin axis
due to the solar torque on the Galilean satellites (which are locked to
the Jupiter's equator plane by this planet's oblateness; Goldreich 1965).
A slight adjustment of satellite or planet positions might be enough
to place Jupiter spin axis precession period
in one side or the other of this resonance, so that
any formation model is apt to be affected. Second, even if the resonance
is crossed the limiting obliquity for adiabatic passage
is significantly smaller than that for the $\nu_{16}$
term. Furthermore, the timescale for adiabatic passage
in the case of the $\nu_{17}$ secular
spin-orbit resonance may be longer ($O(10^7)$ yrs) than the
timescale for gas dissipation by photoevaporation\footnote{Shu \etal (1993)
conclude that EUV from the central star may photoevaporate a T Tauri disk
in $\sim 10^7$ yrs outside of a gravitational radius $r_g \sim 10$ AU.
But there is considerable uncertainty in this
estimate. For instance, Adams \etal
(2004) argue that the disk is likely to be photoevaporated by FUV radiation
from neighboring stars. Furthermore, these authors state
that photoevaporation
is effective to a significantly smaller radius $0.1-0.2 r_g$.}. Third, it is
likely that Uranus and Neptune formed {\it after} Jupiter and
Saturn, once most of gas in the
planetary and circumplanetary disks had already dissipated. Thus, from here on
we focus on $\nu_{16}$ and consider Jupiter and Saturn only.

\section{Secular Perturbation Theory}
\label{sec:SPT}

In a solar system consisting of Jupiter and Saturn it is straightforward
to construct a Laplace-Lagrange secular solution (e.g., Brouwer
and Clemence 1961; Murray and Dermott 1999). For $I << 1$, the orbital
inclination $I$ and the longitude of the ascending node with respect
to the invariant plane $\Omega$ are given by

\begin{equation}
I \sin \Omega = I_1 \sin \gamma_1 + I_2 \sin (f_2 t + \gamma_2),
\end{equation}

\noindent
and

\begin{equation}
I \cos \Omega = I_1 \cos \gamma_1 + I_2 \cos (f_2 t + \gamma_2),
\end{equation}

\noindent
where $f_2$ is the eigenfrequency, $I_1$ and
$I_2$ are eigenvector components, and
$\gamma_1$ and $\gamma_2$ are phases.
If we use parameters
for Jupiter and Saturn as observed, we obtain
$f_2 = - 7.06 \times 10^{-3} \deg$ yr$^{-1}$ and $I_2 = - 6.30 \times 10^{-3}$
(in radians). 
This yields a secular oscillation with period of
$P_{16} \sim 51,000$ yr. These values are not too dissimilar from the
secular solution of the planetary system (e.g., Murray and Dermott 1999).

\begin{figure}
\centerline{\psfig{figure=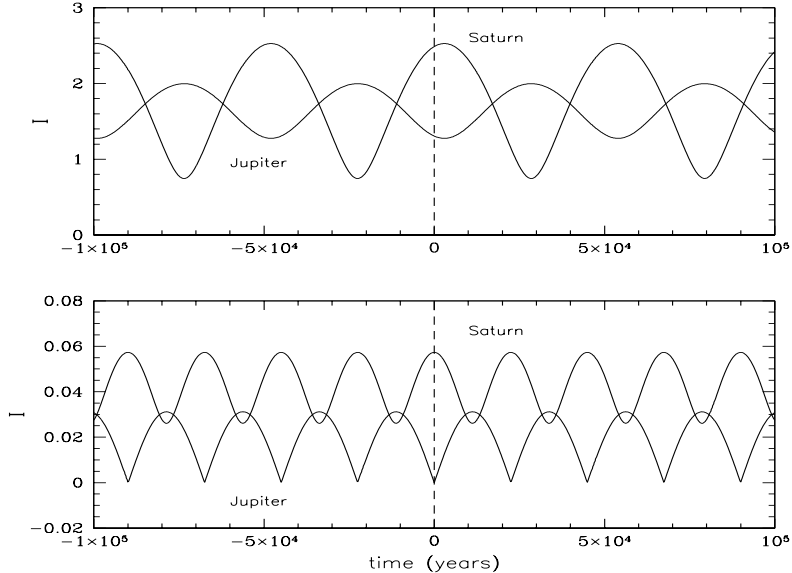,width=4.25in,height=3.25in}}
\caption{Inclinations of Jupiter and Saturn in degrees
derived from a two-body secular perturbation
theory calculated over a time span of $2\times 10^5$ yrs. (a) Case 
centered on 1983 using actual orbital elements (Murray and Dermott 1999). 
(b) Case in which Jupiter has semimajor axis
$a_J = 5.45$ AU, Saturn has $a_S = 8.5$ AU, and their mutual inclination
is $I_{JS} = 10^{-3}$ radians.}
\end{figure}

Let us now consider a time early on before Jupiter and
Saturn had passed through the 1:2 MMR but after most of the planetary
gas disk had dissipated by photoevaporation
in a timescale $10^6-10^7$ yr.
Since the planetary nebula may shield it to some degree, it may be appropriate
to assume that
the much denser subnebula takes longer to dissipate.
If so, at this time both the precession of Jupiter's spin axis and orbital
plane may be significantly faster than they are today. Because
of the scattering on nearby planetesimals (Gomes \etal 2005) and
the dissipation of the nebula, Saturn would 
dominate the precession of Jupiter's orbital
plane.\footnote{Ward (1981) treats
the two-orbit/nebula problem as dispersal takes place.
Here we consider the case when the nebula (but not the subnebula) 
has already been dissipated by some means.}
We take as nominal values for the
orbital parameters of Jupiter and Saturn the starting conditions
in Tsiganis \etal (2005), namely, $a_J = 5.45$ AU, $a_S = 8.50$ AU (a
few tenths of an AU inside the 1:2 MMR) and
$\sin I_{JS} \approx 10^{-3}$, where $I_{JS}$ is the
relative inclination of the
orbital planes of Jupiter and Saturn. With these parameters\footnote{
An estimate for the gap opening timescale may
be obtained using the tidal torque formula (Lin and Papaloizou 1993)
$\tau_{gap} \approx (\Delta/a)^5 P/\mu^2$,
where $\Delta$ is the gap's half-width, $\mu$ is the mass ratio of the
secondary to the primary, and $P$ is the orbital period of the
secondary. Using $\Delta \sim (a_S - a_J)/2 \sim 1.5$ AU,
we find that Jupiter and Saturn
would have cleared the gas disk in between
in a short timescale $t \sim 10^4$ yrs. In particular, this timescale
is shorter than satellite formation timescale $10^4-10^6$
(depending on location; Mosqueira \etal 2001) in MEa,b.} we obtain
$f_2 = -1.60 \times 10^{-2}\deg$ yr$^{-1}$ and $I_2 = -2.72 \times 10^{-4}$
(in radians). As shown in Fig. 1, 
in this case the period of Jupiter's orbital plane 
precession is $P_{16} \sim 23,000$ yrs.

\section{Obliquity Variation}
\label{sec:Obl}
Jupiter's current spin axis precession period is $\sim 4.5 \times 10^5$ yrs
due mostly to the solar torque exerted on the Galilean satellites
(e.g., Ward 1975)\footnote{We obtain a precession period of $4.8 \times
10^5$ yrs using a moment of inertia for Jupiter of $K = 0.26$,
which yields a spin angular momentum $J_p = 4.4 \times 10^{45}$ g cm$^2$/s.
There is some uncertainty in the moment of inertia
$\sim 5 \%$ (Fortney J., pers. comm.)
but this doesn't affect the argument. What is important to note,
however, is that about $33 \%$ of the precession constant $\alpha$ is
contributed by the
torque of the Sun directly on the planet because of its oblateness.
Adding a circumplanetary gas
disk decreases this fraction, which we then ignore.}.
However, a massive circumplanetary
gas disk would result
in a much shorter precession period.
The inner parts of the disk ($r < r_t = \left( 2 M_p/M_\odot 
J_2 R_p^2 a_p^3 \right)^{1/5} \approx 40 R_J$, where
$M_p = M_J$, $a_p = 5.45$ AU, $R_p \approx 2 R_J$, 
and $J_2 \propto 1/R_p \approx 0.008$ are the planetary mass, semi-major
axis, radius and quadrupole gravitational harmonic following
envelope collapse [which may take place in a fast
$10^4-10^5$ yrs timescale, Hubickyj O., pers. comm.], 
and $M_\odot$ and $R_J$ are
the Sun's mass and Jupiter's present radius) would orbit in
the plane of the planet's equator and precess as a unit
with the planet, whereas the outer parts
of the disk ($r > r_t$) would not
(Goldreich 1966).
The spin axis ${\bf \hat{s}}$ then precesses around the orbit normal 
${\bf \hat{n}}$ at
a rate given by (e.g., Tremaine 1991)

\begin{equation}
\frac{d {\bf \hat{s}}}{dt} = \alpha ({\bf \hat{s}} 
\cdot {\bf \hat{n}})({\bf \hat{s}} \times {\bf \hat{n}}),
\end{equation}

\noindent
where ${\bf \hat{s}}$ and ${\bf \hat{n}}$ are unit vectors, and the precession
constant\footnote{In actuality, there should be another
term added to this precession rate due to the torque of the
extended part of the disk $r > r_t$ on the inner part of 
disk. However, the contribution of this extended region
drops rapidly with distance, and in our model the
gas surface density drops-off at a radial location $r_D \lesssim r_t$.} 
is given by

\begin{equation}
\alpha = \frac{3 \pi \Omega_p^2}{2 H} \int_{R_p}^{r_D} \Sigma(a) a^3 da,
\end{equation}

\noindent
where the surface density of
circumplanetary disk $\Sigma(a)$ is assumed to drop-off sharply at
$r_D \sim 2 r_c$, where $r_c = R_H/48 \sim 15 R_J$ 
is the centrifugal radius (MEa),
$\Omega_p$ is the planet's orbital frequency and
the total angular momentum of the precessing system is given by

\begin{equation}
H = J_p + 2 \pi (GM_p)^{1/2} \int_{R_p}^{r_D} 
\Sigma(a) a^{3/2} da,
\end{equation}

\noindent 
where $G$ is the gravitational
constant and $J_p$ is the spin angular momentum of the planet. 
For $\Sigma \propto 1/a$, we can
write $\alpha = M_D \Omega_p^2 r_D^2/(4 H)$ and $H = J_p
+ 2/3 M_D \sqrt{GM_pr_D}$, where $M_D$ is the disk mass.
The precession period is given by $T = 2 \pi/(\alpha \cos \theta)$,
where $\cos \theta = {\bf \hat{s}} \cdot {\bf \hat{n}}$ is the
obliquity.
Given that in our SEMM model
$M_D \sim 10 M_{sats}$, where $M_{sats} \sim 4 \times
10^{26}$ g is the mass of the Galilean satellites\footnote{There
is some ambiguity here because
in the model of MEa,b a significant fraction of the
mass of Callisto is derived from the extended part of the disk, whereas
$M_D$ is the mass of the inner disk out to about Callisto. On the
other hand, Io and Europa should be reconstituted for
unaccreted volatiles.}, the spin of the
planet provides $\sim 90 \%$ of the angular momentum of the system
$H \sim 5 \times 10^{45}$ g cm$^2$/s (see MEa Table 3). 
Using an obliquity of $\theta \sim 3\deg$ and $r_D = 40 R_J$
(which implies a surface density $\Sigma \sim 2 \times 10^4$ g/cm$^2$ at
$15 R_J$ consistent with the value obtained by applying the inviscid
gap-opening criterion to Ganymede in a disk with aspect ratio
$\sim 0.1$ [MEb]),
we obtain a precession period
of $T \sim 4 \times 10^4$ yrs, which is slightly shorter than Jupiter's current
orbital plane secular precession
period $P_{16} \sim 5 \times 10^4$ yrs,
but it is {\it longer} than Jupiter's orbital plane precession period
when Saturn was inside the 1:2 MMR, i.e., $P_{16} \sim 2 \times 10^4$ 
yrs. Hence, in our decaying turbulence
SEMM model it is possible for Jupiter either
to have crossed the secular spin-orbit resonance {\it before} 
the Keplerian disk reached its quiescent phase
(and satellites formed) at a time when the
viscous evolution of the disk was likely driven by Roche-lobe
gas inflow and the resonance passage was non-adiabatic, 
or not to have crossed this
resonance at all. 
However, it may be possible to alter this conclusion by 
choosing different parameters, such as a more massive and extended 
circumplanetary disk.
Thus, next we consider the case of resonance passage.

The limiting obliquity $\theta_{max}$
that could be generated by the obliquity ``kick'' incurred during adiabatic
resonance passage in the non-capture direction ($\dot{\alpha} < 0$) (which
implies resonance crossing as the circumplanetary gas disk is dissipated and
the spin-axis precession period increases) 
is given by (Henrard and Murigande 1987; Ward and Hamilton 2004)

\begin{equation}
\cos \theta_{max} = \frac{2}{\left(1 + \tan^{2/3} |I_2| \right)^{3/2}} - 1,
\end{equation}

\noindent
which yields $\theta_{max} \sim 26\deg$ using the current value
for $|I_2| \sim 0\deg.36$, but a significantly
smaller value of $\theta_{max} \sim 9.1\deg$ using $|I_2| \sim
0\deg.015$ obtained from our nominal, low mutual
inclination case. Furthermore, the minimum time for adiabatic
crossing is (Ward and Hamilton 2004)

\begin{equation}
\tau_{min} \approx P_{16} \left( \frac{\theta}{2 \pi |I_2|} \right)^2.
\end{equation}

\noindent
Taking $\theta \sim 9\deg$, we find
$\sim 2 \times 10^8$ yrs for our nominal case,
which is much longer than the gas dissipation timescale.
This means that at least for the nominal case the crossing must
be non-adiabatic and the final obliquity is rate dependent.
Taking $\theta \sim 3\deg$, we calculate $\tau_{min} \sim 9 \times
10^4$ yrs for current solar-system parameters, but it is
$\tau_{min} \sim 2 \times 10^7$ yrs (here $P_{16} \sim 23,000$ yrs is
about a factor of two shorter than the present value,
but this is more than compensated by the much
smaller value of $|I_2|$) for our nominal case, 
which is comparable to or longer
than the gas dissipation timescale. 
We may estimate the resulting obliquity to be
$\theta \lesssim \tan^{1/3} (|I_2|) \lesssim 4\deg$ 
for our nominal case, which is 
consistent with Jupiter's observed value. 
At any rate,
this would all be irrelevant if the resonance
were not crossed. 

\section{Conclusions}
We have briefly investigated the consequences 
for Jupiter's obliquity of a satellite
formation model in which a massive
(compared to the Galilean satellites yet
enhanced in solids by a factor of $\sim 10$
compared to the solar composition minimum mass model)
subnebula is allowed to dissipate in a long timescale
($10^6-10^7$ yrs) after the dispersal of the nebula itself. 
For the sake of specificity, we have adopted
the model of Tsiganis \etal (2005) in which the
solar system was more compact and regular before
Jupiter and Saturn crossed the 1:2 MMR than it is today.
We find that such a combined scenario (by no means unique)
does not imply the likelihood of
a larger obliquity for Jupiter than is observed. This is both because the
secular $\nu_{16}$ spin-orbit resonance may not be crossed, 
and because the resulting obliquity
may be consistent with Jupiter's value even if it is crossed.
We conclude that
Jupiter's low obliquity is compatible 
with our SEMM satellite formation model (MEa,b) provided
one allows for solar-system conditions early-on unlike those
presently observed.

\vspace{0.2in}
\centerline{\bf Acknowledgements}
This work is supported by grants from PGG
and the National Research Council.

\newpage
\vspace{1in}
\centerline{\bf References} 

\apj{Adams, F. C., Hollenbach, D., Laughlin, G., Gorti, U. 2004}
{Photoevaporation of circumstellar disks due to external far-ultraviolet
radiation in stellar aggregates}{611}{360-379}
\rmp{Balbus, S. A., Hawley, J. F. 1998}{Instability, turbulence, and enhanced 
transport in accretion disks}{70}{1-53}
\refbook{Brouwer, D., Clemence, G. M. 1961}{Methods of Celestial Mechanics.
Academic Press, New York}
\aj{Canup, R. M., Ward, W. R., 2002}{Formation of the Galilean satellites:
conditions of accretion}{124}{3404-3423}
\refbook{Estrada, P. R., Mosqueira, I. 2005}{A gas-poor planetesimal
capture model for the formation of giant planet satellite systems. 
astro-ph/0504649}
\aj{Goldreich, P. 1965}{Inclination of satellite orbits about an oblate
precessing planet}{70}{5-9}
\refbook{Goldreich, P. 1966}{History of the lunar orbit. Rev. Geophys.
{\bf 4}, p.411-439}
\apj{Goldreich, P., Sari, R. 2003}{Eccentricity evolution for planets
in gaseous disks}{585}{1024-1037}
\nature{Gomes, R., Levison, H. F., Tsiganis, K., Morbidelli, A. 2005}
{Origin of the cataclysmic Late Heavy Bombardment period of the
terrestrial planets}{435}{466-469}
\refbook{Hamilton, D., Ward, W. R. 2002}{The obliquities of the giant
planets. American Astronomical Society, DDA Meeting, Vol. 34, p.94}
\aj{Hamilton, D., Ward, W. R. 2004}{Tilting Saturn. II. Numerical
Model}{128}{2510-2517}
\refbook{Hartmann, W. K., Ryder, G., Dones, L., Grinspoon. D. 2000}
{In Origin of the Earth and Moon. (R. Canup and K. Righter,
eds) pp.493-512, Univ. of Arizona Press, Tucson}
\apj{Hawley, J. F., Balbus, S. A., Winters, W. F., 1999}
{Local hydrodynamic stability of accretion disks}
{518}{394-404}
\refbook{Henrard, J., Murigande, C. 1987}{Colombo's top. Celest. Mech.
{\bf 40}, p.345-366}
\refbook{Lin, D. N. C., and J. Papaloizou 1993}{On the tidal interaction
between protostellar disks and companions. In Protostars and Planets III
(E. H. Levy and J. I. Lunine, eds.) pp. 749-836, Univ. of Arizona Press, 
Tucson} 
\refbook{Mosqueira, I., Estrada, P. R., Chambers, J. E., 2000}
{Satellitesimal feeding and the formation of the regular satellites.
American Astronomical Society, 32nd DPS meeting}
\refbook{Mosqueira, I., Estrada, P. R., Cuzzi, J. N., Squyres, S. W., 2001}
{Circumjovian disk clearing after gap-opening and the formation of a partially 
differentiated Callisto. 32nd LPSC meeting, March 12-16, 2001, Houston, Texas, 
no. 1989}
\icarus{Mosqueira, I., Estrada, P. R. 2003a}{Formation of the regular 
satellites of giant planets in an extended gaseous nebula I: subnebula model
and accretion of satellites}{163}{198-231}
\icarus{Mosqueira, I., Estrada, P. R. 2003b}{Formation of the regular 
satellites of giant planets in an extended gaseous nebula II: satellite
migration and survival}{163}{232-255}
\refbook{Murray, C. D., Dermott, S. F. 1999}{Solar System Dynamics.
Cambridge Univ. Press, Cambridge}
\apj{Ogilvie, G. I., Lubow, S. H. 2003}{Saturation of the corotation resonance
in a gaseous disk}{587}{398-406}
\icarus{Shu, F. H., Johnstone, D., Hollenbach, D. 1993}
{Photoevaporation of the solar nebula and the formation of the giant
planets}{106}{92-101}
\icarus{Tremaine, S. 1991}{On the origin of the obliquities of the
outer planets}{89}{85-92}
\nature{Tsiganis, K., Gomes, R., Morbidelli, A., Levison, H. F. 2005}
{Origin of the orbital architecture of the giant planets of the
Solar System}{435}{459-461}
\aj{Ward, W. R. 1975}{Tidal friction and generalized Cassini's laws in
the solar system}{80}{64-70}
\icarus{Ward, W. R. 1981}{Solar nebula dispersal and the stability
of the planetary system}{47}{234-264}
\refbook{Ward, W. R. 2003}{Constraints on the Galilean protosatellite
disk from Jupiter's obliquity. AGU, Fall Meeting 2003}
\aj{Ward, W. R., Hamilton, D. 2004}{Tilting Saturn. I. Analytic Model}
{128}{2501-2509}

\end{document}